\documentclass[conference]{IEEEtran}
\IEEEoverridecommandlockouts

\usepackage{upquote}
\usepackage{lscape}
\usepackage{ifpdf}
\usepackage{lscape}
\usepackage{cite}
\ifCLASSINFOpdf
\usepackage[pdftex]{graphicx}

\else
  \usepackage{epstopdf}
  \usepackage[dvips]{graphicx}
  \usepackage[demo]{graphicx}
    \usepackage{subcaption}
\fi
\usepackage{amsmath}
\usepackage{multirow, makecell}
\usepackage{algorithmic}
\usepackage{array}
\ifCLASSOPTIONcompsoc
  \usepackage[caption=false,font=normalsize,labelfont=sf,textfont=sf]{subfig}
\else
  \usepackage[caption=false,font=footnotesize]{subfig}
\fi

\usepackage{url}
\begin{document}

\title{Speech Imagery Classification using Length-Wise Training based on Deep Learning\\

\thanks{20xx IEEE. Personal use of this material is permitted. Permission
from IEEE must be obtained for all other uses, in any current or future media, including reprinting/republishing this material for advertising or promotional purposes, creating new collective works, for resale or redistribution to servers or lists, or reuse of any copyrighted component of this work in other works. This work was partly supported by Institute of Information \& Communications Technology Planning \& Evaluation (IITP) grant funded by the Korea government (MSIT) (No. 2017-0-00432, Development of Non-Invasive Integrated BCI SW Platform to Control Home Appliances and External Devices by User’s Thought via AR/VR Interface; No. 2017-0-00451, Development of BCI based Brain and Cognitive Computing Technology for Recognizing User’s Intentions using Deep Learning; No. 2019-0-00079, Artificial Intelligence Graduate School Program (Korea University)).}

}

\author{\IEEEauthorblockN{Byeong-Hoo Lee}
\IEEEauthorblockA{\textit{Dept. Brain and Cognitive Engineering}\\
\textit{Korea University} \\
Seoul, Republic of Korea \\
bh\_lee@korea.ac.kr}\\

\IEEEauthorblockN{Do-Yeun Lee}
\IEEEauthorblockA{\textit{Dept. Brain and Cognitive Engineering}\\
\textit{Korea University} \\
Seoul, Republic of Korea \\
doyeun\_lee@korea.ac.kr}\\

\and

\IEEEauthorblockN{Byeong-Hee Kwon}
\IEEEauthorblockA{\textit{Dept. Brain and Cognitive Engineering}\\
\textit{Korea University} \\
Seoul, Republic of Korea \\
bh\_kwon@korea.ac.kr}\\

\IEEEauthorblockN{Ji-Hoon Jeong}
\IEEEauthorblockA{\textit{Dept. Brain and Cognitive Engineering}\\
\textit{Korea University} \\
Seoul, Republic of Korea \\
jh\_jeong@korea.ac.kr}\\

}
\maketitle

\begin{abstract}
Brain-computer interface uses brain signals to control external devices without actual control behavior. Recently, speech imagery has been studied for direct communication using language. Speech imagery uses brain signals generated when the user imagines speech. Unlike motor imagery, speech imagery still has unknown characteristics. Additionally, electroencephalography has intricate and non-stationary properties resulting in insufficient decoding performance. In addition, speech imagery is difficult to utilize spatial features. In this study, we designed length-wise training that allows the model to classify a series of a small number of words. In addition, we proposed hierarchical convolutional neural network structure and loss function to maximize the training strategy. The proposed method showed competitive performance in speech imagery classification. Hence, we demonstrated that the length of the word is a clue at improving classification performance.

\end{abstract}

\begin{small}
\textbf{\textit{Keywords-brain-computer interface; electroencephalogram; speech imagery; convoulutional neural network}}\\
\end{small}

\IEEEpeerreviewmaketitle

\section{Introduction}
Brain-computer interface (BCI) uses brain signals to control the external devices without actual control behavior and explore the statements of users \cite{C1, MRCP, lee2015subject,lee2017network,  bci, Kwak3, BinHe, park2016movement}. Unlike invasive BCI, non-invasive BCI does not require brain surgery \cite{chen2016high}. One of the mainly used brain signals in non-invasive BCI is electroencephalography (EEG) \cite{highertemporal, chen2017extraction}. Three major EEG-based BCI paradigms are movement-related cortical potentials (MRCP) \cite{MRCP}, event-related potentials (ERP) \cite{ERP}, and motor imagery (MI) \cite{MI}. MRCP reflects the user's voluntary movement a few seconds prior \cite{karimi2017detection} and many studies used MRCP for exoskeleton walking experiments \cite{roboticarm, Kwak2, sawicki2020exoskeleton}. An ERP is a potential generated by direct brain responses such as P300 and N200 from a stimulus such as a target cognitive \cite{ERPspeller, won2017motion, lee2018high}. MI is the most studied endogenous paradigm on account of the neurophysiological origin \cite{zhang2020motor, suk2014predicting, C2}. It is generated over the motor cortex area when the user imagines muscle movements \cite{chholak2019visual, suk2011subject}. In this paradigm, signals are represented cortex event-related desynchronization/synchronization (ERD/ERS) patterns \cite{ERD}. Recently, speech imagery has been studied for direct communication using language. Similar to MI, speech imagery uses brain signal generated when the user imagines specific words. MI has known crucial characteristics such as the contra laterality and functional areas of the brain (spatial features) for classification, but speech imagery has the unknown. Therefore, numerous studies have investigated the characteristics of speech imagery using various methods \cite{Nguyen, chi2011eeg}.

\begin{figure*}[!t]
  \centerline{\includegraphics[width =\textwidth ]{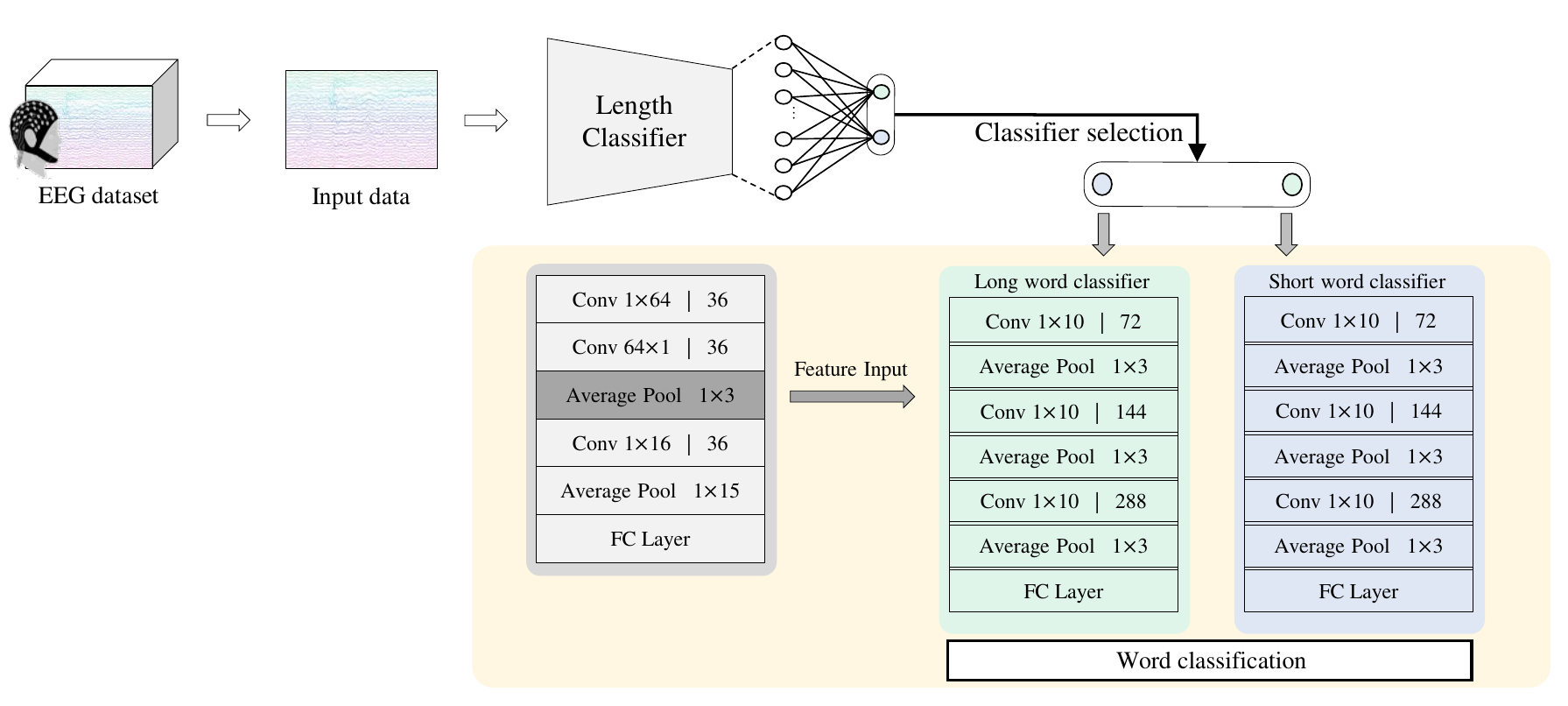}}
  \caption{Overall flow chart of the proposed method. Length classifier first predicts word length and selects corresponding word classifier. Selected classifier performs word classification using feature input from length classifier. During training, word classifiers use the feature input regardless of prediction of length classifier. Therefore, they can use more training samples.}
\end{figure*}

In the endogenous paradigm, decoding user intention is a challenging task because EEG signals have a low signal-to-noise ratio and non-stationary properties \cite{nonstationary}. To overcome this limitation, MI paradigms generally consist of preparation sessions or motor execution sessions to active brain regions. However, the speech imagery paradigm does not require any preparation sessions because it is easier to perform and repeat than MI. This property leads to increased consistency in the dataset \cite{Nguyen} that is a great advantage in classification. Therefore, recent advanced methods have investigated suggesting novel feature extraction techniques based on machine learning and deep learning. DaSalla $et \ al.$\cite{DaSalla} proposed single-trial classification method based on common spatial pattern (CSP). Torres-Garcia $et\ al.$ \cite{Torres-Garcia} implemented fuzzy inference based channel selection method for imagined speech classification. Ngyuyen $et\ al.$ \cite{Nguyen} collected the speech imagery dataset and proposed vector machine classifiers based on Riemannian manifold. However, it is necessary to increase classification performance for practical uses of speech imagery. There is still plenty of room to increase classification performance.

In this paper, we proposed length-wise training to improve classification performance using length classifier and two word classifiers. Through the length-wise training, the proposed method trains word length classifier and word classifiers at the same time. The length classifier predicts word length and selects a word classifier. We designed hyper-label named short and long word, one syllable and more than two-syllable, for word length classifier training. Word classifiers use feature input produced by word length classifier. We designed loss function to take into account uncertainty of prediction of word length classifier. Therefore, each classifier classifies only a small number of words that can improve classification performance.

Hence, our main contributions represented in three folds: 1) We investigated the effect of the length of the word (syllable) on the classification. 2) We designed length-wise training and relevance training strategy to improve classification performance that enables each classifier classifies a small number of classes. 3) The proposed method achieved remarkable improvement in speech imagery classification performance, and we have proved that length-wise training can improve multi-class speech imagery classification.

\begin{table}[!t]
{\normalsize
\caption{Architecture of The Proposed Method}
\renewcommand{\arraystretch}{1.2}
\resizebox{\columnwidth}{!}{%
\begin{tabular}{llll} \Xhline{4\arrayrulewidth}
                         & Length Classifier        & Long-word Classifier           & Short-word Classifier            \\\hline
Input                    & Raw EEG (1, 1, 64, 512)          & Feature input (1, 36, 1, 150)  & Feature input (1, 36, 1, 150)  \\\hline
\multirow{10}{*}{Layers} & Conv2D (1, 64), 36 filter     & Conv2D (1, 10), 72 filter     & Conv2D (1, 10), 72 filter     \\
                         & Conv2D (64, 1) 36 filter     & AvgPool (1, 3, stride: 1, 3) & AvgPool (1, 3, stride: 1, 3) \\
                         & AvgPool (1, 3, stride: 1, 3)  & Conv2D (1, 10), 144 filter    & Conv2D (1, 10), 144 filter    \\
                         & Conv2D (1, 16), 36 filter      & AvgPool (1, 3, stride: 1, 3) & AvgPool (1, 3, stride: 1, 3) \\
                         & AvgPool (1, 15, stride: 1, 15) & Conv2D (1, 10), 288    & Conv2D (1, 10), 288 filter    \\
                         & Fully-Connected Layer         & AvgPool (1, 3, stride: 1, 3) & AvgPool (1, 3, stride: 1, 3) \\
                         & Softmax                       & Fully-Connected Layer        &  Fully-Connected Layer \\
                         &                               & Softmax                      & Softmax \\
                   \\\hline
Activation               & ELU                           & ELU                          & ELU                          \\\hline
Optimizer                & AdamW                          & AdamW                         & AdamW                         \\\hline
Loss function            & Cross entropy                 & Cross entropy                & Cross entropy                \\\hline
\end{tabular}}}
\end{table}

\section{Methods}

\subsection{Overall Framework}
We designed a length classifier and two word classifiers for efficient speech imagery classification. In the following section, we describe the details of convolutional neural network (CNN) architectures and training strategy including training process and loss function. 

\subsubsection{Length Classifier}
The length classifier selects a word classifier based on the predicted word length. We assigned hyper-label according to the syllable of words, length classifier starts training using the hyper-label (long or short word). Therefore, length classifier classifies word length according to hyper-label and selects a word classifier. The length classifier sends the features, extracted after receptive field generation, to word classifiers. The temporal and spatial convolution layer sets the receptive field from 4 Hz to remove ocular artifacts and reduce the channel dimension to a single dimension. Hence, length classifier and word classifiers share the same receptive field and channel dimension.

\subsubsection{Word Classifiers}
Two word classifiers predict long and short words respectively. In general, a single CNN predicts certain words among all words at once. On the other hand, word classifiers predict only the words assigned. Hence, the proposed method performs a series of a small number of classification. Word classifiers consist of three pairs of convolution-pooling layers for feature extraction. AdamW optimizer with weight decay (0.01) \cite{adamW}, ELU activation function \cite{ELU} and cross-entropy loss function \cite{shore1980axiomatic} were used for training.

\subsection{Training Strategy}
For efficient training, we designed a length-wise training strategy. As mentioned above, length classifier predicts the word length and selects corresponding word classifier and the selected word classifier classifies the word. However, both word classifiers classify the word regardless of prediction of length classifier during training. One word classifier trained using the correctly assigned word but the other classifier trained incorrectly assigned word. Therefore, word classifiers trained using more training samples than when using only correctly assigned words. The details of design choices are described in Table I.

The prediction of length classifier has uncertainty because the classifier does not always predict the length accurately. To consider this uncertainty, we aggregated outputs of length classifier with a loss function. Since the outputs are produced through the softmax layer, we can handle them as probabilities. Therefore, the loss function consisted of multiplication of probability and training loss. The details of the loss function are described in the ``Loss Function" section. 

\subsection{Loss function}
We designed three CNN architectures, thus they produce three loss values. Total loss is a summation of three loss values represented as:

\begin{equation}
loss = L_w + L_s+ L_l
\end{equation}
where ${L_w}$, ${L_s}$, and ${L_l}$ are loss of length classifier, short and long word classifier respectively. Length classifier produces two probabilities, probability of short word $p_{s}$ and long word $p_{l}$, and we aggregated them to cross-entropy loss function defined as:

\begin{equation}
L_{s} = -p_{o.s}\sum_{1}^{M}y_{o.s}\log{\bar{y}_{o.s}}
\end{equation}
\begin{equation}
L_{l} = -p_{o.l}\sum_{1}^{N}y_{o.l}\log{\bar{y}_{o.l}}
\end{equation}
where $M$ and $N$ are the number of short and long words and $o$ denote trial (training sample). Probability of $o$ is multiplied to loss function to reduce the loss value according to length classifier outputs. Through this modification, the proposed method considers the uncertainty of the word length classifier during the training.

\section{Results and Discussions}
We used the BCI competition 2020 track 3 dataset ($http://brain.korea.ac.kr/bci2021/competition.php$) that consists of 300 trials of the training dataset (60 trials per class) and 50 trials of validation dataset (10 trials per class). Test dataset is not available at this point, we split the training dataset into training and validation dataset. Therefore we used validation dataset as a test dataset. We applied 5-fold cross-validation, training dataset consists of 48 trials per class and the validation dataset consists of 12 trials per class. The performance was calculated as the average of the folds. Additionally, we applied a cropped decoding technique using 1 s window and about 125 ms of stride. Models produce multiple outputs, one output per crop, and make a prediction using averaged output \cite{deepconvnet}. We assigned ``$hello$", ``$help\ me$" and ``$thank \ you$" classes as long words, more than two-syllable words, and ``$stop$" and ``$yes$" as short words. Therefore, a long word classifier classifies three classes, on the other hand, short word classifier classifies two words.

\begin{table}[]
{\normalsize
\caption{Classification Results of The Proposed Method}
\renewcommand{\arraystretch}{1.2}
{\begin{center}

\begin{tabular}{c|c|c|c}\hline
Model         & Mean (std)    & Median & Max - Min \\ \hline
Shallow \cite{deepconvnet} & 53.47 (7.27) & 52     & 64  - 36  \\ \hline
Deep  \cite{deepconvnet} & 56.13 (9.52) & 54     & 70  - 40  \\ \hline
EEGNet  \cite{EEGNET} & 58.27 (9.62) & \textbf{62}     & \textbf{74}  - 38  \\ \hline
Proposed & \textbf{59.47 (6.86)} & \textbf{62}     & 70  - \textbf{48} \\ \hline
\end{tabular}\end{center}}}
\end{table}

\begin{figure*}[!t]
  \centerline{\includegraphics[scale = 0.8 ]{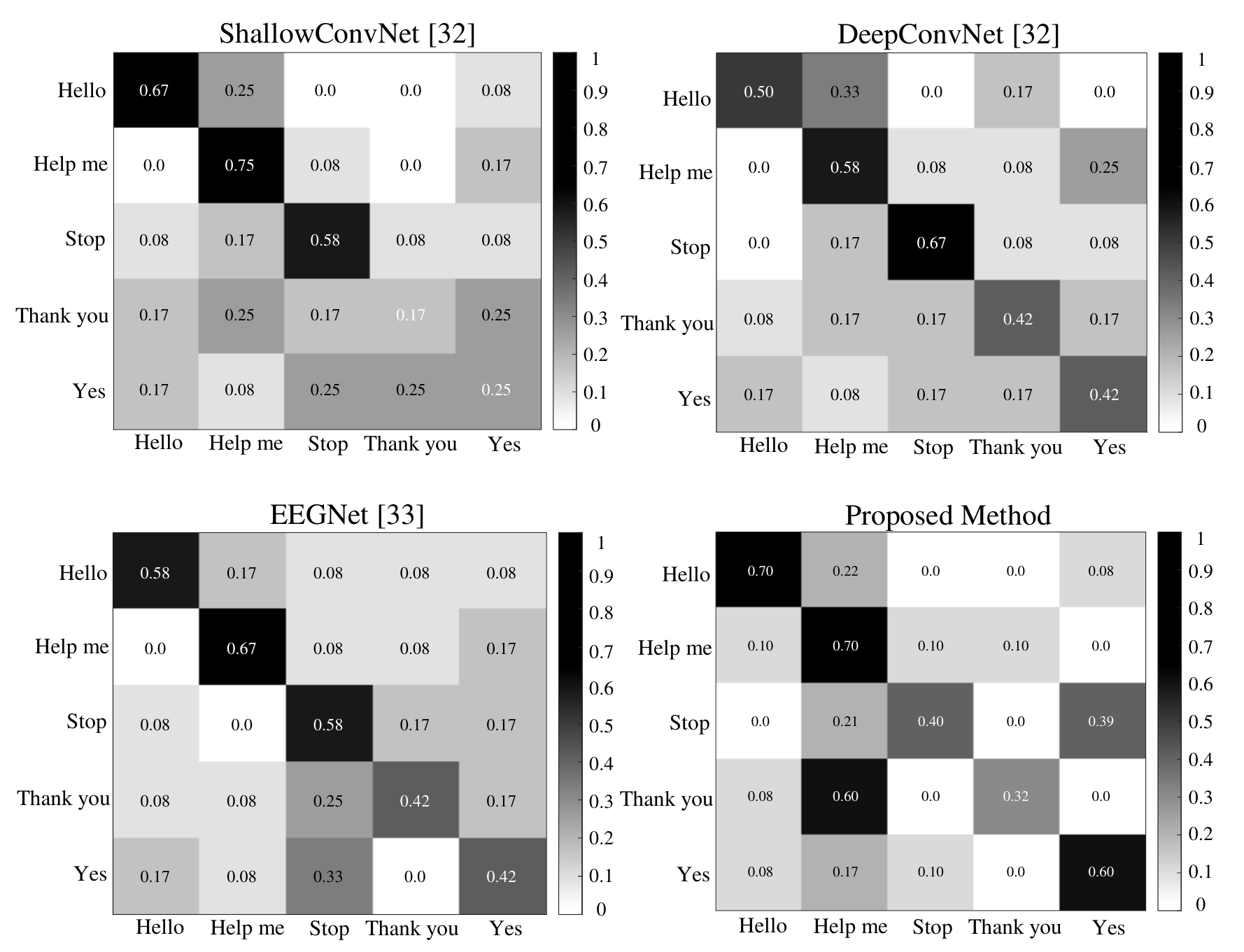}}
  \caption{Confusion matrices of the models. There is different tendency between proposed method and existing models. The proposed model tended to confuse between words of the same length. Especially it misclassified the ``$thank \ you$'' as ``$help \ me$''. On the other hand, the others tended to confuse the words ``$thank \ you$'' and ``$yes$''.}
\end{figure*}

Unlike MI classification, focusing on band power features, shows inefficiency in speech imagery. Therefore, it is not advantageous to design a model shallower. ShallowConvNet consisted of three layers showed the lowest accuracies in mean and median values. However, the difference between the minimum and maximum accuracy of ShallowConvNet was the smallest among the models. DeepConvNet showed about 3 percent higher accuracy than ShallowConvNet across the evaluation indexes. Frequency-specific spatial filters allow model to extract meaningful features leading to performance improvement. EEGNet that consisted of a depthwise and separable convolution layer achieved the highest median and maximum accuracy. Length-wise training shows the efficiency in speech imagery that allows the model to classify the length first. The proposed method achieved the highest accuracy where the highest mean, median and minimum accuracy. Therefore, it has been proved that length-wise training is efficient for speech imagery classification. We present the classification results in Table II. 

We conducted an additional analysis using the confusion matrix presented in Fig. 2. Through the analysis, we confirmed that there is a different classification tendency between the proposed method and the other methods. The proposed method tends to confuse the same length words. In particular, it misclassified ``$thank \ you$'' with the word ``$help \ me$'' (0.60). In addition, the proposed method misclassified the word ``$stop$'' with ``$yes$'' about 0.39 points. On the other hand, the other models show low accuracies in ``$thank \ you$'' and ``$yes$''. ShallowConvNet shows the highest accuracy in ``$hello$'', ``$help \ me$'' and ``$stop$'' classification but shows the lowest accuracy in ``$thank \ you$'' and ``$yes$'' classification. DeepConvNet misclassified ``$hello$'' with ``$help \ me$'' (0.33) and ``$help \ me$'' with ``$yes$'' (0.25). EEGNet confused `$thank \ you$'' and ``$yes$'' with ``$hello$''.

\section{Conclusion and Future Works}
\label{sec:print}
In this paper, we proposed the length-wise training that trains word length classification before word classification using word length classifier. Additionally, we designed the loss function to take account of the uncertainty of length classifier output. Speech imagery classification is challenging because it has unknown characteristics to use for classification. Furthermore, improving classification accuracy is essential for the practical use of speech imagery. We demonstrated that length-wise training achieved meaningful improvement in speech imagery classification. Therefore, we confirmed that the length of the word is a clue at improving classification performance. We will improve the classification performance between the same length words. Furthermore, the proposed method could be applied to help intuitively control external devices such as a robotic arm with a high degree of freedom. Therefore, the proposed method will help improve the life quality of people with movement disabilities.

\section{Acknowledgement}
The author would like to thank J.-H. Cho for help with the discussion of the data analysis.\\
\bibliographystyle{./IEEEtran}
\bibliography{./IEEEexample}

\end{document}